\documentclass{epl}
\usepackage{amsmath}
\usepackage{amssymb}

\newcommand{\PFlowlevel}[1]{(TMTSF)$_2$#1}
\newcommand{\PF}{\PFlowlevel{PF$_6$}}           
\newcommand{\X}{\PFlowlevel{X}}                 
\newcommand{\myrefeq}[1]{Eq.~(\ref{#1})}
\newcommand{\myreffig}[1]{Fig.~\ref{#1}}
\newcommand{\cs}[1]{\textsf{#1}}
\newcommand{\QP}{quasiparticle}
\newcommand{\QPE}{quasiparticle energy}
\newcommand{\ie}{{\it i.e.}}

\newcommand{\Omcminv}{\un{(\Omega\,\text{cm})^{-1}}}

\newcommand{\parl}[2]{\mv{#1}\|\mv{#2}}
\newcommand{\Tst}{T^\star}                          
\newcommand{\Tsdw}{T_{\text{SDW}}}
\newcommand{\mdeg}{^\circ}
\newcommand{\MR}{\Delta\rho/\rho_0}
\newcommand{\sncn}[1]{\sin^2\theta+\gamma_{#1}\cos^2\theta}
\newcommand{\ssncn}[1]{\sqrt{\sncn{#1}}}
\newcommand{\veps}{\varepsilon_0}

\newcommand{\mv}[1]{\mathbf{#1}}                
\newcommand{\vek}[1]{$\mv{#1}$}                 
\newcommand{\mvk}{\mv{k}}
\newcommand{\vk}{\vek{k}}

\newcommand{\mvB}{\mv{B}}   \newcommand{\vB}{$\mvB$}
\newcommand{\mva}{\mv{a}}   \newcommand{\va}{$\mva$}
\newcommand{\mvb}{\mv{b'}}  \newcommand{\vb}{$\mvb$}
\newcommand{\mvc}{\mv{c^*}} \newcommand{\vc}{$\mvc$}
\newcommand{\mvj}{\mv{j}}   \newcommand{\vj}{$\mvj$}

\newcommand{\mja}{\parl{j}{a}}          \newcommand{\ja}{$\mja$}
\newcommand{\mjb}{\parl{j}{b'}}         \newcommand{\jb}{$\mjb$}
\newcommand{\mjc}{\parl{j}{c^\star}}    \newcommand{\jc}{$\mjc$}

\newcommand{\mBa}{\parl{B}{a}}          \newcommand{\Ba}{$\mBa$}
\newcommand{\mBb}{\parl{B}{b'}}         \newcommand{\Bb}{$\mBb$}
\newcommand{\mBc}{\parl{B}{c^\star}}    \newcommand{\Bc}{$\mBc$}

\newcommand{\plane}[2]{#1\text{--}#2}
\newcommand{\macplane}{\plane{\mva}{\mvc}}      \newcommand{\acplane}{$\macplane$}
\newcommand{\mbcplane}{\plane{\mvb}{\mvc}}      \newcommand{\bcplane}{$\mbcplane$}
\newcommand{\mabplane}{\plane{\mva}{\mvb}}      \newcommand{\abplane}{$\mabplane$}

\newcommand{\mBab}{\parl{\mvB}{(\mabplane)}}    \newcommand{\Bab}{$\mBab$}
\newcommand{\mBbc}{\parl{\mvB}{(\mbcplane)}}    \newcommand{\Bbc}{$\mBbc$}
\newcommand{\mBac}{\parl{\mvB}{(\macplane)}}    \newcommand{\Bac}{$\mBac$}

\newcommand{\Tfourlowlevel}[1]{$\Tst#1 4\un{K}$}
\newcommand{\Tfour}{\Tfourlowlevel{\approx}}        

\title{Magnetoresistance in the SDW state of \PF\ above \Tfour; Novel effect due to the
Landau quantization} \shorttitle{Magnetoresistance in the SDW state of \ldots}

\author{B. Korin-Hamzi\'{c}\inst{1}\thanks{E-mail: \email{bhamzic@ifs.hr}}
             \and M. Basleti\'{c}\inst{2} \and K. Maki\inst{3}}
\shortauthor{B. Korin-Hamzi\'{c} \etal}

\institute{
  \inst{1} Institute of Physics, POB 304, HR-10001 Zagreb, Croatia\\
  \inst{2} Department of Physics, Faculty of Science, POB 331, HR-10001 Zagreb, Croatia\\
  \inst{3} Department of Physics, University of Southern California, Los Angeles CA 90089-0484, USA
           and
           Max-Planck Institute for the Physics of Complex Systems, N\"{o}thnitzer Str.38, D-01187 Dresden,
           Germany
}




\begin{document}

\maketitle

\begin{abstract}
Magnetoresistance in the spin-density wave (SDW) state of \PF\ is known to exhibit a rich variety of the
angular dependencies when a magnetic field \vB\ is rotated in the \bcplane, \abplane\ and \acplane\
planes. In the presence of a magnetic field the \QP\ spectrum in the SDW with imperfect nesting is
quantized. In such a case the minimum \QPE\ depends both on the magnetic field strength $|B|$ and the
angle $\theta$ between the field and the crystal direction \va, \vb\ or \vc. This approach describes
rather satisfactory the magnetoresistance above \Tfour.
\end{abstract}

\section{Introduction}
Since the discovery of the superconductivity in \PF\ in 1979~\cite{JeromeJP80} the Bechgaard salts or
the highly anisotropic organic conductors \X\ (where TMTSF is tetramethyltetraselenafulvalene and X is
anion PF$_6$, AsF$_6$, ClO$_4$ \ldots) are well known because of a variety of physical phenomena related
to their low dimensionality. Their rich phase diagram exhibits various low temperature phases under
pressure and/or in magnetic field among which the spin density wave (SDW), field induced SDW (FISDW)
with quantum Hall effect, and unconventional superconductivity are the most intriguing
\cite{IshiguroBook98,LeePRL02}. The quasi-one-dimensionality (1D) is a consequence of the crystal
structure, where the TMTSF molecules are stacked in columns in the \va\ direction (along which the
highest conductivity occurs), and the resulting anisotropy in conductivity is commonly taken to be
$\sigma_a : \sigma_b : \sigma_c \approx t_a^2 : t_b^2 : t_c^2\approx 10^5:10^3:1$, where $t_i$ are the
tight binding transfer integrals. These materials are open-orbit metals with the Fermi surface (FS)
consisting of a pair of weakly modulated sheets in the \vek{b} and \vek{c} directions.

The SDW ground state of the organic conductors \X\ has been the subject of considerable experimental and
theoretical investigations in the last twenty years~\cite{IshiguroBook98}. It is caused by the almost
perfect nesting of the two sheets of the FS, due to the instability of the (1D) electron gas with strong
electron-electron interactions. The opening of the energy gap ($\Delta$) in the conduction band leads to
the semiconducting transport properties. The external pressure increases the transverse coupling and
above about 8~kbars the SDW ground state is suppressed in favor of superconductivity.

One of the most intriguing features of the \X\ is their anomalous behaviour in a magnetic field. There
are several angular effects of magnetoresistance known in the literature, like the variety of
resonance-like features, known as the geometrical resonance or the Lebed resonance (or the magic angle
effect)\cite{LebedJETP86}, the Danner-Chaikin oscillation~\cite{DannerPRL95} and the third angular
effect~\cite{OsadaPRL96}, and all of them are mostly connected with the metallic phase of \X. Only few
measurements of the magnetoresistance anisotropy, as far as we know, have been performed in the SDW
ground state at ambient pressure \cite{UlmetJPL85,BasleticEPL93,KorinHamzicEPL98,KorinHamzicJP99}.

\PF\ is one of the most investigated \X\ compound. It is metallic down to $\Tsdw \approx 12\un{K}$ where
the metal-semiconductor transition into SDW ground state occurs and below which the resistance displays
an activated behavior. It is known that SDW in \PF\ undergoes another transition at $\Tst \approx
\Tsdw/3$ (at $3.5-4\un{K}$ at ambient pressure)~\cite{LebedJETP86,DannerPRL95}, but the nature of the
possible subphases remains controversial. We have recently shown
\cite{KorinHamzicEPL98,KorinHamzicJP99,BasleticPRB02} that the temperature dependent magnetoresistance
anisotropy changes abruptly below 4~K, indicating a possible phase transition at $\Tst \approx 4\un{K}$.

In this paper we shall concentrate ourselves to the magnetoresistance (MR) behavior above $\Tst$, and we
propose that it can be understood in terms of the Landau quantization of the \QP\ spectrum in a magnetic
field $B$, where the imperfect nesting plays the crucial role. The experimental MR data of \PF\ at
$T=4.2\un{K}$ will be compared with our new theoretical model. The discussion of the magnetoresistance
below $\Tst$, in terms of unconventional SDW (or USDW), will be presented elsewhere
\cite{BasleticPRB02,KorinHamzicIJMPB01}.

\section{Experiment}
The measurements were done in magnetic fields up to 5~T with directions of the current along the
different crystal axis, and for different orientations of the applied magnetic field. A rotating sample
holder enabled the sample rotation around a chosen axis over a range of $190\mdeg$. The experimental MR
data are for \vc\ (\jc), \vb\ (\jb) and \va\ (\ja) axis and for different orientations of magnetic
field. The single crystals used come all from the same batch. Their \va\ direction is the highest
conductivity direction, the \vb\ direction (with intermediate conductivity) is perpendicular to \va\ in
the \abplane\ plane, and \vc\ direction (with the lowest conductivity) is perpendicular to the \abplane\
plane (and $\plane{\mva}{\mv{b}}$). The room temperature conductivity values are: $\sigma_a =
500\Omcminv$, $\sigma_b = 20\Omcminv$ and $\sigma_c = 1/35\Omcminv$.

\begin{figure}[b]
\onefigure[width=12cm]{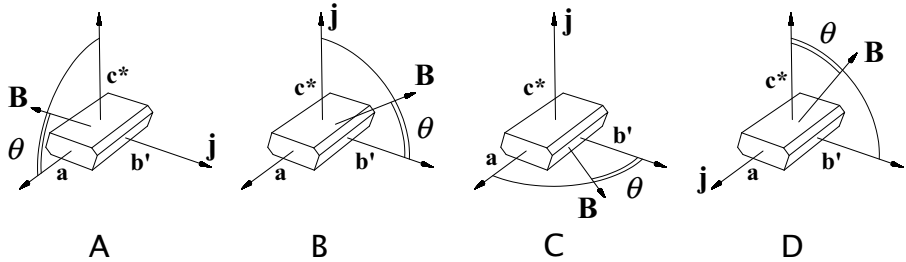} \caption{Four configurations (\cs{A}, \cs{B}, \cs{C}\ and \cs{D}) of
the current \vj\ and magnetic field \vB\ direction. (See text for a detailed explanation.)}
\label{fig:config}
\end{figure}

The MR (defined as $\MR = [\rho(B)-\rho(0)]/\rho(0)$) was measured in various four probe arrangements on
samples cut from the long crystals. In the case of $\rho_a$ (\ja), $\rho_b$ (\jb) and $\rho_c$ (\jc),
two pairs of the contacts were placed on the opposite \bcplane, \acplane\ and \abplane\ surfaces,
respectively.

The four configurations that will be analyzed in this work are shown on
Fig 1\cs{A}, \cs{B}, \cs{C} and \cs{D}:\\
{\it i)} \myreffig{fig:config}\cs{A} shows the case when the current direction is along the \vb\ axis
and the magnetic field is rotated in the \acplane\ plane (\jb, \Bac) perpendicular to the current
direction. $\theta$ is the angle between \vB\ and the \va\ axis, \ie\ $\theta=0$ for \Ba\ and
$\theta=90\mdeg$ for
\Bc.\\
{\it ii)} \myreffig{fig:config}\cs{B} shows the case when the current direction is along the \vc\ axis
and the magnetic field is rotated in the \bcplane\ plane (\jc, \Bbc). $\theta$ is the angle between \vB\
and the
\vb\ axis, \ie\ $\theta=0$ for \Bb\ and $\theta=90\mdeg$ for \Bc.\\
{\it iii)} \myreffig{fig:config}\cs{C} shows the case when the current direction is along the \vc\ axis
and the magnetic field is rotated in the \abplane\ plane (\jc, \Bab) perpendicular to the current
direction. $\theta$ is the angle between \vB\ and the \vb\ axis, \ie\ $\theta=0$ for \Bb\ and
$\theta=90\mdeg$ for \Ba.\\
{\it iv)} \myreffig{fig:config}\cs{D} shows the case when the current direction is along the \va\ axis
and the magnetic field is rotated in the \bcplane\ plane (\ja, \Bbc) perpendicular to the current
direction. $\theta$ is the angle between \vB\ and the \vc\ axis, \ie\ $\theta=0$ for \Bc\ and
$\theta=90\mdeg$ for \Bb.

\section{Model, Results and Discussion}
\begin{figure}
\onefigure[height=4.5cm]{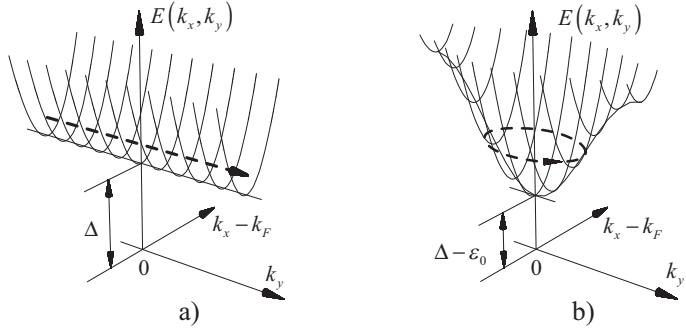} \caption{$E(k_x,k_y)$ in SDW with a) perfect nesting, and b)
imperfect nesting.} \label{fig:landscape}
\end{figure}

If we limit ourselves to SDW above \Tfour, it is well established that the \QPE\ is given
by~\cite{KorinHamzicEPL98,HuangPRB90}:
\begin{equation}\label{eq:QPEnergy}
    E(\mvk) = \sqrt{\eta^2 + \Delta^2} -  \veps\cos 2bk_y\,,
\end{equation}
where $\eta = \sqrt{ v_a^2 \left(k_x-k_F\right)^2 + v_c^2 k_z^2 }$ is the \QPE\ in the normal state
($v_a$ and $v_c$ are Fermi velocities in \va\ and \vc\ direction), $\Delta\approx 34\un{K}$ is the order
parameter for SDW and $\veps\approx 13\un{K}$ is the parameter characterizing the imperfect nesting
\cite{KorinHamzicEPL98}. In a presence of a magnetic field $B$ the \QP\ orbit is quantized. This can be
readily seen from the \QPE\ landscape as shown in \myreffig{fig:landscape}.
In \myreffig{fig:landscape}a we show the \QPE\ for SDW with perfect nesting. Quasiparticle energy has
the minima at $k_x=\pm k_F$, which is independent of $k_y$ and consequently, the \QP\ orbit is open. On
the other hand, for SDW with imperfect nesting, there are minima for \QPE\ at $k_x=\pm k_F$ and $k_y =
\pm \pi/2b$ (see \myreffig{fig:landscape}b). Therefore, in the presence of a magnetic field, \QP\ will
circle around these minima, \ie\ closed orbits appear and they will be quantized.

We expand the \QPE\ in \myrefeq{eq:QPEnergy} for small $(k_x-k_F)^2$ and $k_y^2$. In the presence of a
magnetic field within different planes (\myreffig{fig:config}) the minimum \QPE\ (\ie\ the energy gap)
associated with the lowest Landau level is given by:
\begin{eqnarray}\label{eq:EnergyGapAbove}
E(B,\theta) & = & \Delta - \veps + \sqrt{\frac{\veps}{\Delta}}\,v_a b e B\ssncn{2}\,,
                \text{\ \ for}\mBbc \\
            & = & \Delta - \veps + \frac{v_a v_c b e}{2 \Delta} B\ssncn{3}\,,
                \text{\ \ for}\mBab \\
            & = & \Delta - \veps + \sqrt{\frac{\veps}{\Delta}}\,v_a b e B|\sin\theta|\,,
                \text{\ \ for }\mBac\,,
\end{eqnarray}
where $\gamma_2 = \gamma_3^{-1} = \left( v_c/2b \right)^2 / \veps\Delta$. In general, the \QPE\ gap
increases linearly with $|B|$ and depends on angle $\theta$ (as defined in \myreffig{fig:config}). For
$B=0$ the resistance is given as $\rho_i \propto \exp \left[ \beta E(0,0) \right]$. For $B \neq 0$ and
for $\omega_c\tau > 1$ ($\omega_c$ -- cyclotron frequency; $\tau$ -- scattering rate; we also assume
that the \QP\ scattering rate is independent of \vk) we may write down the magnetoresistance as:
\begin{eqnarray} \label{eq:MRzzBbc}
\rho_{zz}(T,B) & = &
        \exp    \left\{
                    \beta(\Delta-\veps) \left[ 1+A_2 B \ssncn{2} \right]
                \right\} \nonumber \\
   & & \times \left(  1+C_2 B \ssncn{2} \right)\,, \text{\ \ for}\mBbc
\end{eqnarray}
\begin{eqnarray} \label{eq:MRzzBab}
\rho_{zz}(T,B) & = &
        \exp    \left\{
                    \beta(\Delta-\veps) \left[ 1+A_3 B \ssncn{3} \right]
                \right\} \nonumber \\
   & & \times \left(  1+C_3 B \ssncn{3} \right)\,, \text{\ \ for}\mBab\,,
\end{eqnarray}
with:
\begin{equation}\label{eq:AAzz}
    A_2 = \sqrt{ \frac{\veps}{\Delta} } \, \frac{v_a b e} {\Delta-\veps}\,,\quad
     A_3 = \frac{v_a v_c e}{2\Delta} \, \frac{1}{\Delta-\veps} \,.
\end{equation}

\begin{figure}
\twofigures[width=6.8cm,clip]{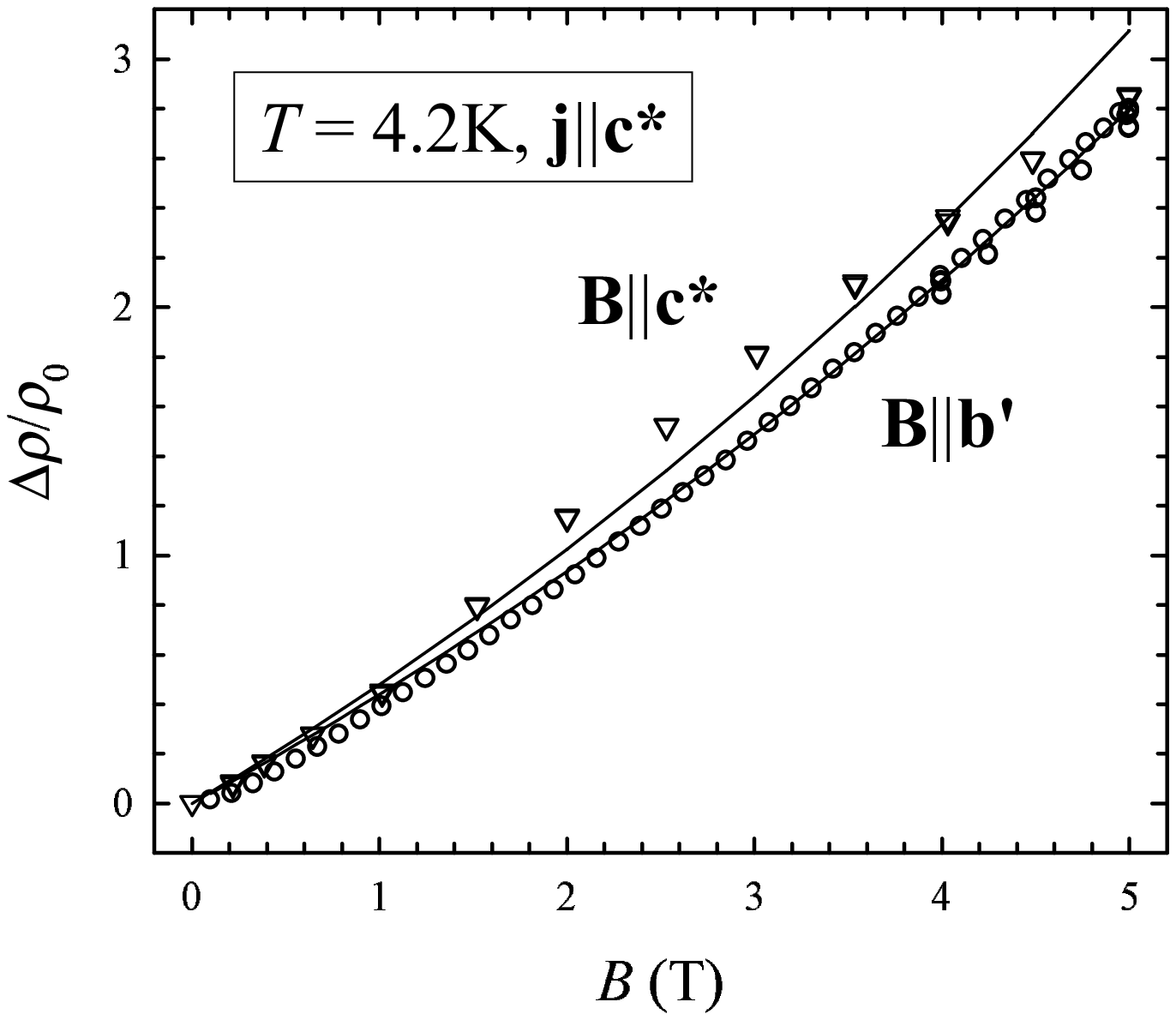}{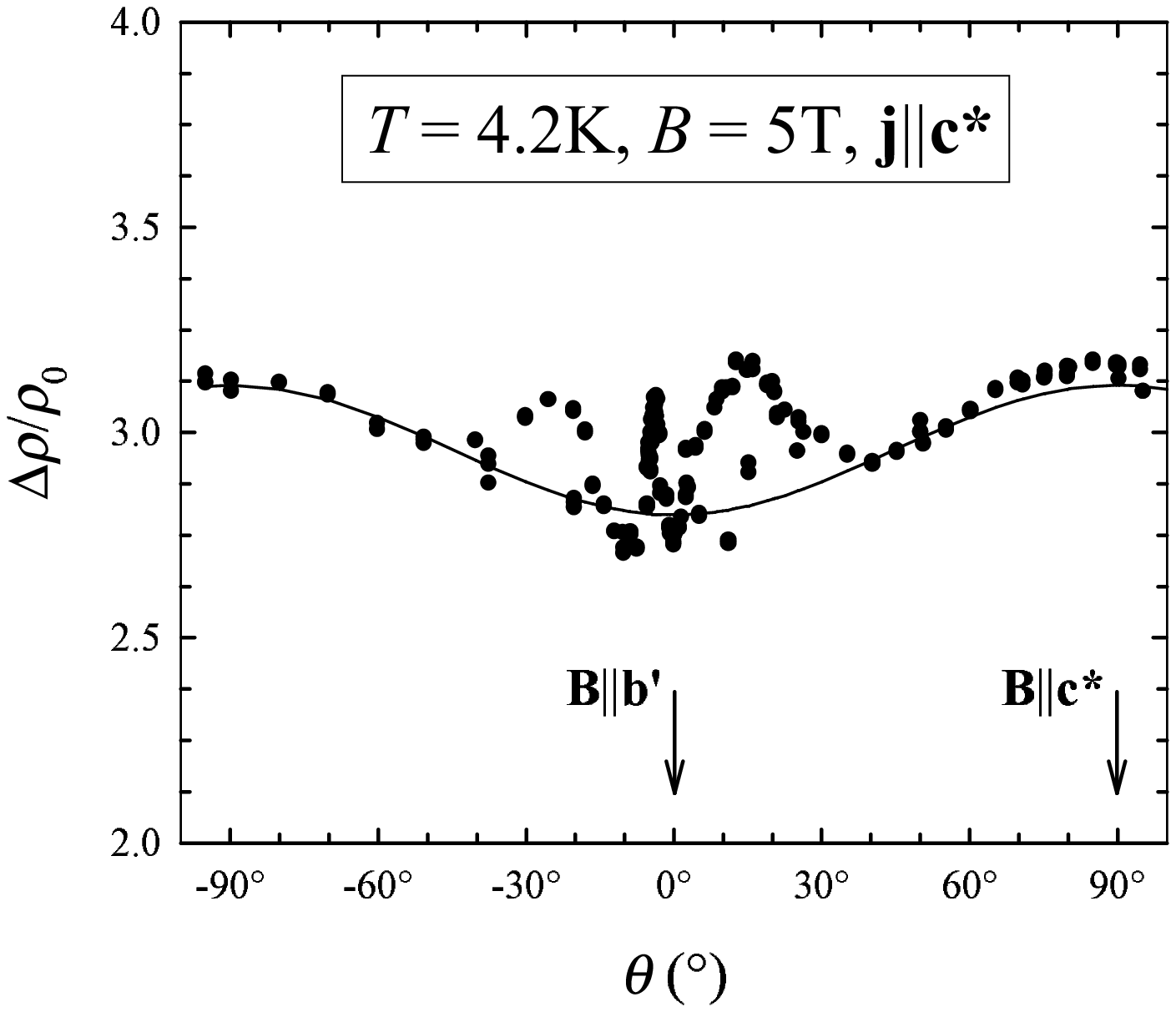}

\caption{Magnetic field dependence of $\MR$ at 4.2~K for \jc, \Bb\ and \Bc\ (see
\myreffig{fig:config}\cs{B}). Solid lines are fits to the theory (see text).} \label{fig:MRzzBbc}

\caption{Angular dependence of $\MR$ at 4.2~K and $B=5$~T, for \jc, \vB\ in \bcplane\ plane (see
\myreffig{fig:config}\cs{B}). Solid lines are fits to the theory (see text).} \label{fig:MRzzThbc}
\end{figure}

\begin{figure}
\twofigures[width=6.8cm,clip]{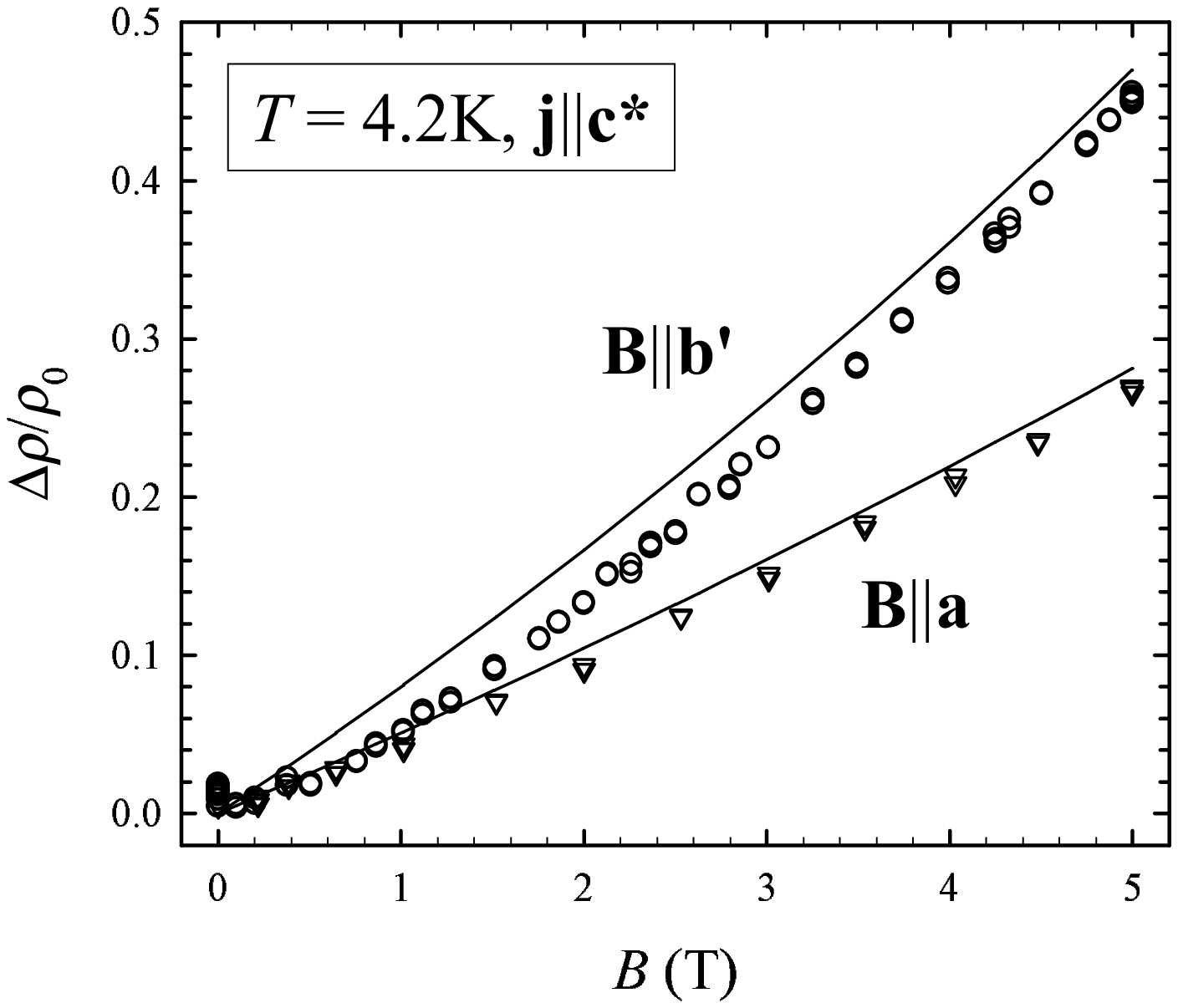}{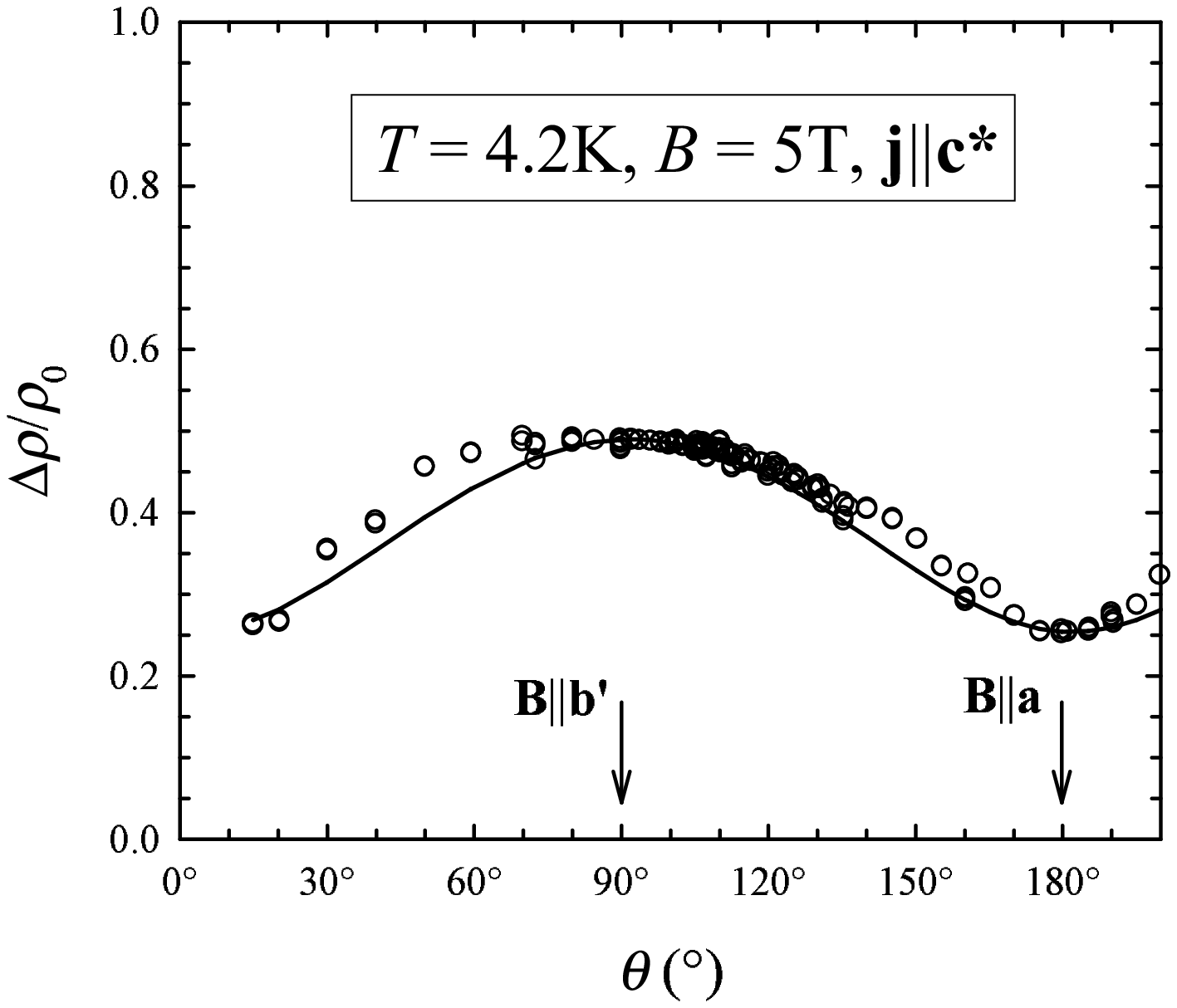}

\caption{Magnetic field dependence of $\MR$ at 4.2~K for \jc, \Ba\ and \Bb\ (see
\myreffig{fig:config}\cs{C}). Solid lines are fits to the theory (see text).} \label{fig:MRzzBab}

\caption{Angular dependence of $\MR$ at 4.2~K and $B=5$~T, for \jc, \vB\ in \abplane\ plane (see
\myreffig{fig:config}\cs{C}). Solid lines are fits to the theory (see text).} \label{fig:MRzzThab}
\end{figure}

We compare now our experimental data with the above equations. Figs.~\ref{fig:MRzzBbc} and
\ref{fig:MRzzThbc} show magnetic field dependence of MR (\jc, \Bc\ and \Bb) and the angular dependence
of MR (\jc, \Bbc) at 4.2~K, respectively. The solid lines are fits to \myrefeq{eq:MRzzBbc}. The
analogous results for \jc\ but different magnetic field rotation (\Bab) is given on
Figs.~\ref{fig:MRzzBab} and \ref{fig:MRzzThab}. The solid lines are fits to \myrefeq{eq:MRzzBab}.

\begin{figure}
\twofigures[width=6.8cm,clip]{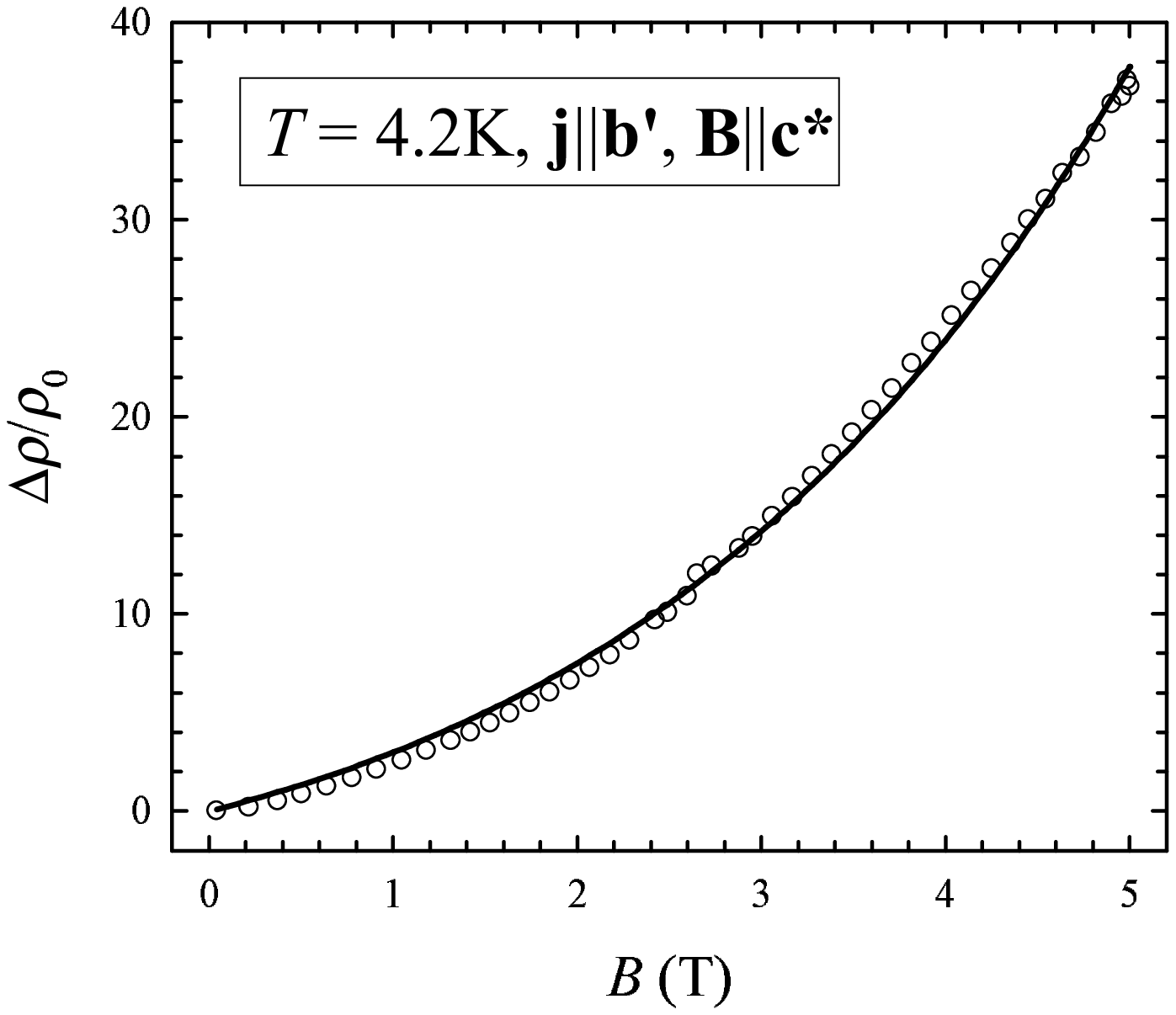}{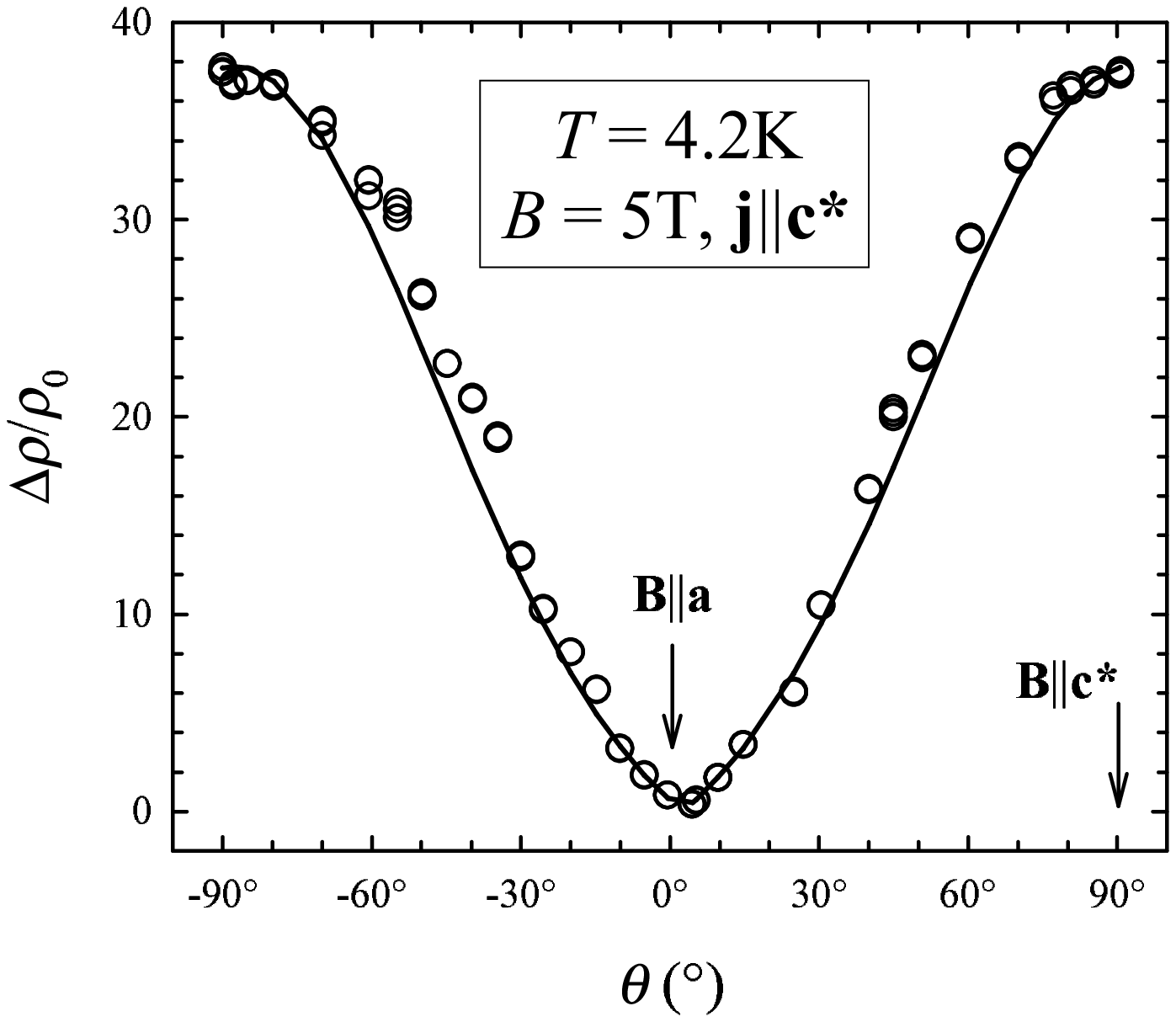}

\caption{Magnetic field dependence of $\MR$ at 4.2~K for \jb, \Bc\ (see \myreffig{fig:config}\cs{A}).
Solid lines are fits to the theory (see text).} \label{fig:MRyyBac}

\caption{Angular dependence of $\MR$ at 4.2~K and $B=5$~T, for \jb, \vB\ in \acplane\ plane (see
\myreffig{fig:config}\cs{A}). Solid lines are fits to the theory (see text).} \label{fig:MRyyThac}
\end{figure}

On the other hand, for \jb\ and \Bac\ we obtain:
\begin{equation} \label{eq:MRyy}
\frac{\Delta\rho_{yy}(T,B)}{\rho_{yy}(T,0)}  =
        \exp    \left\{ \beta(\Delta-\veps) A_1 B |\sin\theta| \right\}
    \left(  1+ C_1 B |\sin\theta| \right)\,,
\end{equation}
where $A_1 = \sqrt{ \veps/\Delta } \, v_a b e / (\Delta-\veps)$. We give the excess magnetoresistance
rather than the MR itself, as in this configuration the \QPE\ gap for $B=0$ is only
8~K~\cite{KorinHamzicEPL98}, instead of all other cases where $\Delta-\veps = 21\un{K}$. The
experimental results for \jb\ and \vB\ in \acplane\ plane are shown on Figs.~\ref{fig:MRyyBac} and
\ref{fig:MRyyThac}. The solid lines are fits to \myrefeq{eq:MRyy}.

It is evident that the present model (Eq.~(\ref{eq:MRzzBbc}--\ref{eq:MRyy})) describes both the $B$ and
$\theta$ dependence of MR rather well (fits on Figs.~\ref{fig:MRzzBbc}--\ref{fig:MRyyThac} are from good
to excellent). The fitting procedure yielded $\Delta-\veps=21\un{K}$, $A_2 = 0.014\un{T^{-1}}$,
$\gamma_2 = 0.85$, $C_2 = 0.38\un{T^{-1}}$, $A_3 = 0.00905\un{T^{-1}}$, $\gamma_3 = 3.1$, $C_3 = 0$,
$A_1 = 0.048\un{T^{-1}}$, $C_1 = 2.14\un{T^{-1}}$.


First, from $A_2$ and $\gamma_2$ we can extract the \va\ axis coherence length $\xi_a = v_a/\Delta =
120$~\AA\ and $v_c/v_a = 7.33\times 10^{-2}$. The ratio $v_c/v_a$ appears to be somewhat larger than the
one previously determined ($v_c/v_a = 1.7\times 10^{-2}$ \cite{IshiguroBook98}) but the difference is
minor. Furthermore, $A_3$ gives $\xi_a = 100$~\AA, where we used our $v_c/v_a$ value. We can, therefore,
conclude that $\xi_a$ is consistent in these two configurations. On the other hand, $\gamma_3$ gives
$\xi_a = 5.4$~\AA\ (the value that appears to be too small) and $A_1$ gives $\xi_a = 410$~\AA\ (which is
somewhat larger, but the difference of a factor 3.5 may be acceptable).

\begin{figure}
\twofigures[width=6.8cm,clip]{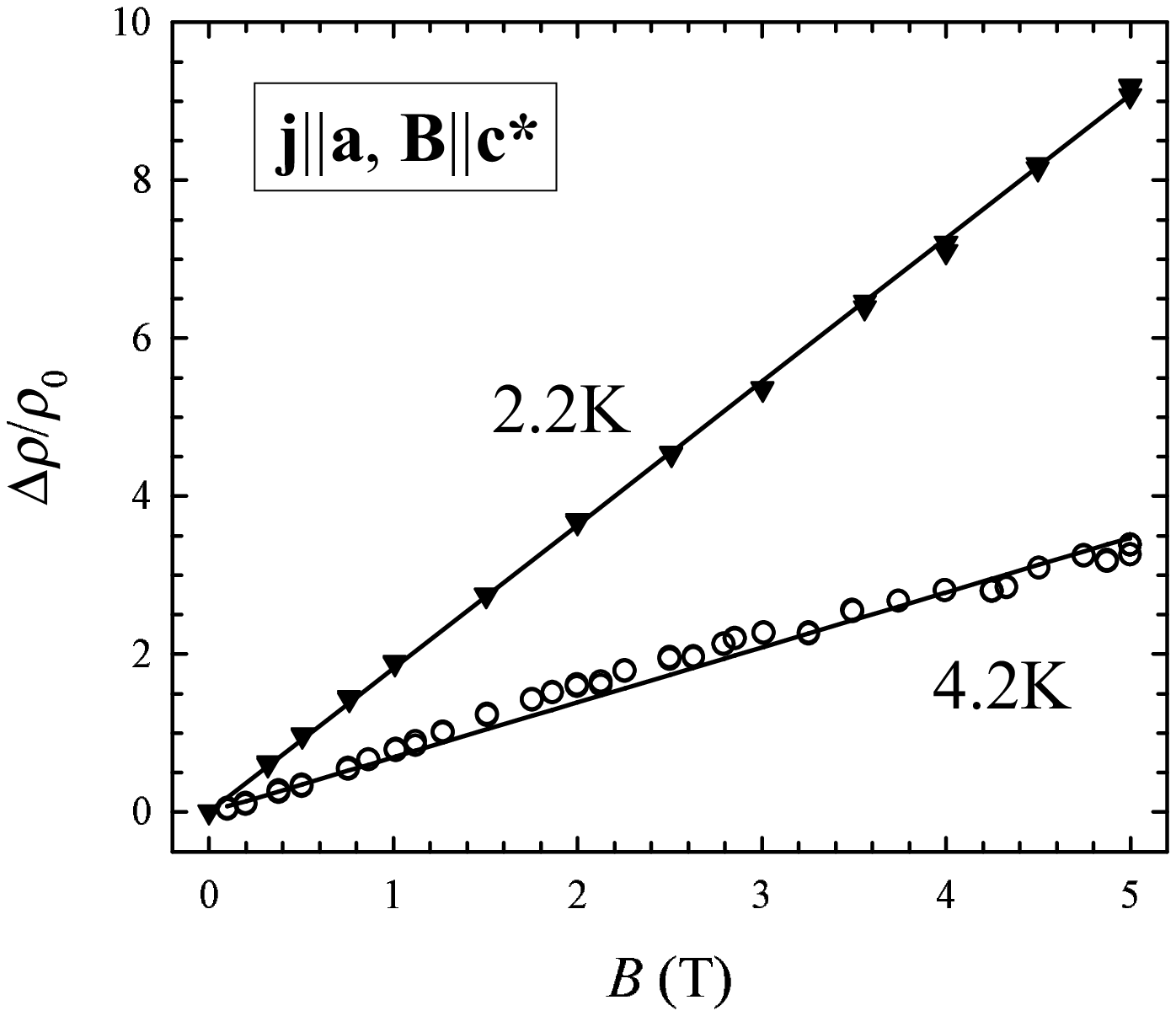}{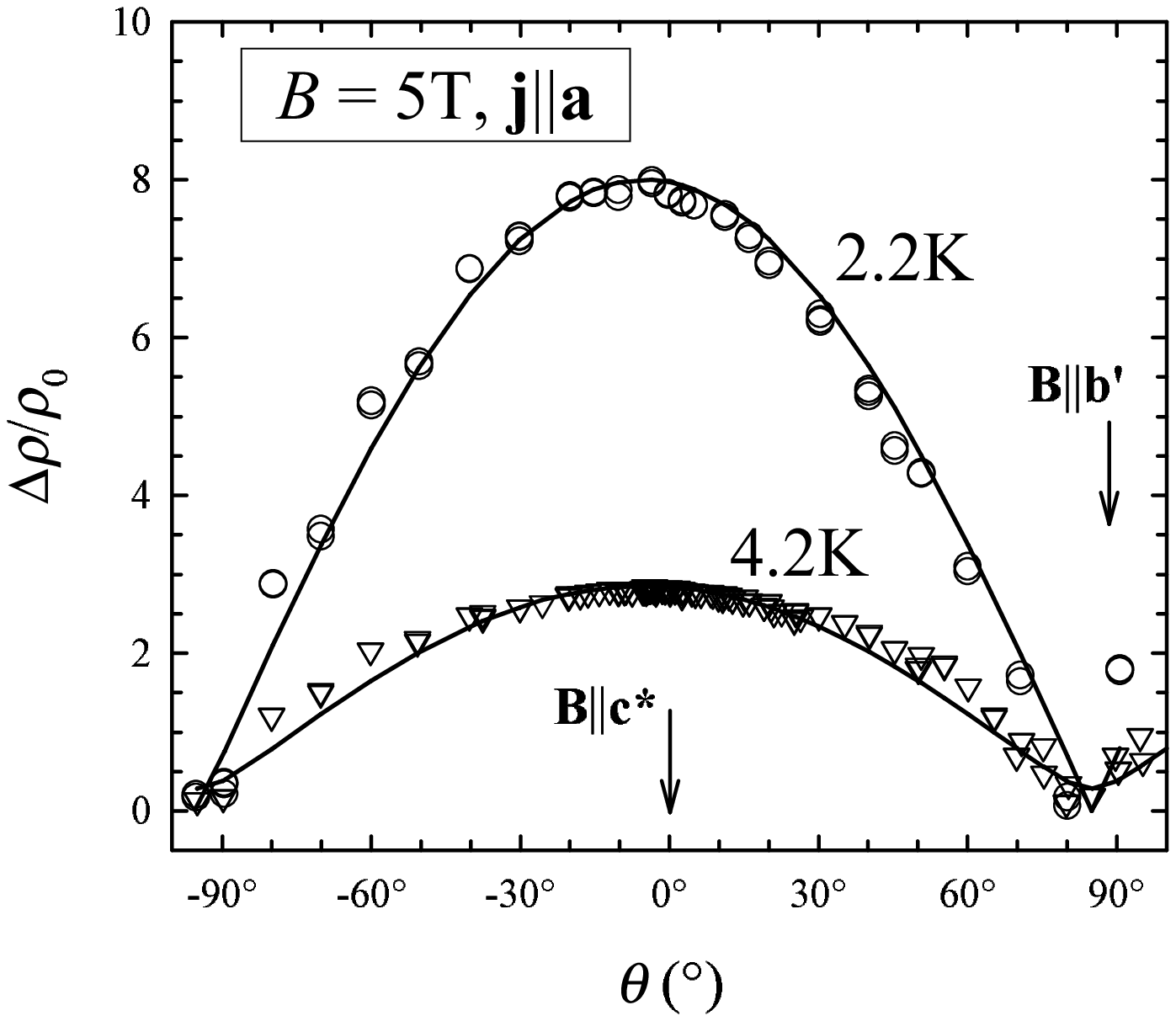}

\caption{Magnetic field dependence of $\MR$ at 4.2~K and 2.2~K, for \ja, \Bc\ (see
\myreffig{fig:config}\cs{D}). Solid lines are fits to the theory (see text).} \label{fig:MRxxBbc}

\caption{Angular dependence of $\MR$ at 4.2~K and 2.2~K, for $B=5$~T and \ja, \vB\ in \bcplane\ plane
(see \myreffig{fig:config}\cs{D}). Solid lines are fits to the theory (see text).} \label{fig:MRxxThbc}
\end{figure}

Finally, we comment the result for the highest conductivity direction \va, \ie\ $\rho_{xx}(T,B)$ for
\vB\ in the \bcplane\ plane, perpendicular to the current direction. The experimental results for the
magnetic field dependence of MR (\ja, \Bc\ at 4.2~K and 2.2~K) and the angular dependence of MR at 4.2~K
and 2.2~K are shown on Figs.~\ref{fig:MRxxBbc} and \ref{fig:MRxxThbc}. We notice that the angular
dependencies of MR for $T=4.2\un{K}$ and $2.2\un{K}$ look very similar (see \myreffig{fig:MRxxThbc}).
Namely, the fitted $\rho_{xx}(T,B)$ for both $T=4.2\un{K}$ and $2.2\un{K}$ (Figs.~\ref{fig:MRxxBbc} and
\ref{fig:MRxxThbc}, solid lines), are well accounted for by:
\begin{equation}\label{eq:MRxx}
    \rho_{xx}(B,\theta) \propto 1 + C_1 B|\cos\theta|\,,
\end{equation}
where $\theta$ is the angle $\measuredangle(\mvB,\mvc)$. Here $C_1 = 0.55\un{T^{-1}}$ and
$1.6\un{T^{-1}}$ for $T=4.2\un{K}$ and $2.2\un{K}$, respectively.

The linear $B$ dependence of $\rho_{xx}$ for \Bbc\ in SDW state of \PF\ (\myreffig{fig:MRxxBbc}) is well
known \cite{IshiguroBook98}, though it is not understood. Therefore, we may conclude that
$\rho_{xx}(T,B)$ is not sensitive to the \QP\ spectrum we are considering here.

\section{Conclusion}
In summary, we have derived the expression of the magnetoresistance based on the Landau quantization of
the \QP\ orbit in SDW with imperfect nesting. At the qualitative level these expressions give excellent
description of experimental magnetoresistance results. From fitting procedure, we can deduce the
physical content like $\xi_a$ and $v_c/v_a$, which both look very reasonable. The origin of a few cases
where $\xi_a$ has not a consistent value remains to be clarified. In any case, we may conclude that the
Landau quantization of the \QPE\ describes most of salient features of the angular dependence of the
magnetoresistance in SDW in \PF\ above $\Tst\approx 4\un{K}$.

A parallel study of the magnetoresistance for $T<\Tst$ based on possible USDW in addition to the already
existing SDW, will be reported elsewhere~\cite{KorinHamzicIJMPB01,BasleticPRB02}.

\acknowledgments

This experimental work was performed on samples prepared by K.~Bechgaard. We acknowledge useful
discussion with A.~Hamzi\'{c} and S.~Tomi\'{c}.

\end{document}